\documentclass[aps,prb,twocolumn,showpacs,superscriptaddress,amsmath,amssymb]{revtex4}

\usepackage{graphicx}
\usepackage{color,colordvi}
\usepackage{epstopdf}
\usepackage{dcolumn}
\usepackage{bm}
\usepackage{psfrag}

\begin{document}
\title{Nickel assisted healing of defective graphene}
\author{S. Karoui}
\affiliation{Laboratoire d'Etudes des Microstructures, ONERA-CNRS, BP 72, 92322 Ch\^atillon Cedex, France}
\author{H. Amara}
\affiliation{Laboratoire d'Etudes des Microstructures, ONERA-CNRS, BP 72, 92322 Ch\^atillon Cedex, France}
\author{C. Bichara}
\affiliation{Centre Interdisciplinaire de Nanoscience de Marseille,  (CINaM), CNRS and Aix Marseille University,
 Campus de Luminy, case 913, 13288 Marseille Cedex 09, France}
\author{F. Ducastelle}
\affiliation{Laboratoire d'Etudes des Microstructures, ONERA-CNRS, BP 72, 92322 Ch\^atillon Cedex, France}
\date{\today}
\begin{abstract}

The healing of graphene grown from a metallic substrate is investigated using tight-binding Monte Carlo simulations. At temperatures (ranging from 1000 to 2500 K), an isolated graphene sheet can anneal a large number of defects suggesting that their healing are thermally activated. We show that in presence of a nickel substrate we obtain a perfect graphene layer. The nickel-carbon chemical bonds keep breaking and reforming  around defected carbon zones, providing a direct interaction, necessary for the healing. Thus, the action of Ni atoms is found to play a key role in the reconstruction of the graphene sheet by annealing defects. 

\end{abstract}


\maketitle

\section{ Introduction}

Carbon structures --- graphene and nanotubes in particular --- are the focus of substantial research activity, due to their promising electrical, mechanical and optical properties.~\cite{Saito98, Loiseau06} Particular attention has been paid recently to single layers of graphite, \textit{i.e.} to  graphene, and to its novel electronic properties.~\cite{Geim07, Castro09} In many applications, optimal performance requires the control of its structural properties, which remains a significant difficulty. It is known for a long time that Catalytic Chemical Vapor Deposition (CCVD) of hydrocarbons on reactive nickel or transition-metal-carbide surfaces can produce thin graphitic layers.~\cite{Shelton74}  However, for the most part, the large amounts of carbon atoms absorbed on nickel foils formed thick graphite crystals instead of graphene films. Few layer graphene is now commonly produced by CCVD at temperatures ranging between 350 and 1000$^\circ$C, using metallic substrates (such as Co, Ni, Ir, Ru) that catalyze the decomposition of the carbon-bearing gas precursors and make the growth of carbon nanostructures.~\cite{Coraux08, Yu08, Sutter08, Loginova09, Kwon09, Sutter09, Reina09} Large-area graphene films of the order of centimeters are now grown on copper substrates possible.~\cite{Li09}

Despite this considerable progress in synthesis procedures, the details of the microscopic mechanisms involved in the growth of carbon structures are still lacking. Atomic scale investigations are far from being feasible experimentally, whereas computer simulations allow such investigations. We have recently developed a tight-binding (TB) model for nickel and carbon that uses Monte Carlo simulations in the grand canonical ensemble (GCMC) to study the formation of graphene from a metallic substrate.~\cite{Amara01, Amara02}  The GCMC method works in an open system and can simulate the growth of a carbon structure under conditions of fixed temperature and carbon chemical potential, closely reproducing the experimental conditions where the carbon chemical potential is fixed by the catalytic decomposition of the carbon-bearing gas. Hence, we were able to understand and explain the nucleation of graphene on Ni (111). The Ni surface catalyses the nucleation process: carbon atoms are confined in its surroundings until a critical surface concentration is reached. Carbon chains appear first, interacting weakly with the surface and becoming more and more stable; then nucleation of graphene takes place. The energetics of the carbon--metal interaction, a limited carbon solubility, and a tendency for expelling the $sp^{2}$ carbon structures formed at the catalyst surface are therefore important. 

However, for a large range of chemical potentials, we observe the presence of defects in the grown $sp^{2}$ structure. Likewise, in the case of carbon nanotubes grown from a metallic nanoparticle, whether the employed method is empirical,\cite{Ribas09} or semi-empirical,\cite{Page09, Amara03} all the final configurations are plagued by a high concentration of atomic-scale defects. These include, but are not limited to, heptagon-pentagon topological defects, adatoms, and atomic vacancies. Furthermore, the estimated experimental growth rates are in the nanometer to micrometer per second range.~\cite{Yoshida08, Picher09} Such scales are largely beyond the capabilities of numerical simulations, meaning that only some elementary steps can be studied, or that the growth conditions imposed in the simulations are orders of magnitude too fast, leading to very defective $sp^{2}$ carbon structures. In this context, several groups have sought to characterize the crystallization of carbon network using numerical simulations.~\cite{Page09, Lee05, Zhang07}

In the present work, we investigate the healing processes of defective carbon structures at atomic scale. We have developed an improved Tight-Binding Monte Carlo (TBMC) code that allows us to perform systematic and long Monte Carlo runs. We use it  to study the evolution at finite temperature of graphitic structures previously grown by grand canonical Monte Carlo (GCMC) simulations. We also investigate the role played by the catalyst metal in the healing process of graphitic structures.

\section{Method}

We have developed and carefully tested a model for Ni and C, in a tight-binding framework.~\cite{Amara01, Amara02}  The total energy is a sum of local terms: an empirical repulsive term and a band structure contribution including $s$ and $p$ electrons of C, and $d$ electrons of Ni. Local densities of electronic states are calculated using the recursion method. Only the first four continued fraction coefficients ($a_{n}, b_{n}$) are calculated exactly. The continued fraction is then expanded to the $N^{th}$ level using constant coefficients equal to $a_{2}$ and $b_{2}$ (here $N = 40$). This approach provides us already with a good description of the angular contributions to the energy and leads to a relatively fast scheme. This atomic interaction model is then implemented in a Monte Carlo code, based on the Metropolis algorithm, which allows for structure relaxation. To go further, we have improved the Monte Carlo code efficiency in canonical ensemble by modifying the algorithms.~\cite{Jan10} Switching from the standard recursion algorithm to a direct calculation of the moments is possible and more efficient when only four moments are considered. Indeed in a Monte Carlo algorithm, we can calculate only the moments contribution that have changed at each step. A second point concerns the way to calculate the local densities of states, by using an analytic integration of the continued fraction expansion.~\cite{Allan84} This option is faster but is less stable and more difficult to implement.  By combining both developments, we have obtained a faster code, that makes the exploration of longer time scales and more extensive investigations possible. More details will be given elsewhere. We have studied the influence of temperature variations on defective graphene by using this approach. The idea is to start from graphene sheets containing defects and submit them to high temperatures up to 2500 K,  which is a typical temperature used in graphitization processes.\cite{Setton02} The system is then able to overcome high energy barriers and reach new equilibrium states corresponding to less defective carbon structures.

\section{Stone-Wales defect}

As a first step towards a complete understanding of the healing process in graphene, we study the influence of temperature variations on the Stone-Wales defect. The Stone-Wales (SW) defect is a 90$^{\circ}$ rotation of a C-C bond in the hexagonal network with respect to the midpoint of the bond. This leads to the formation of two pentagons and two heptagons, replacing four hexagons. This defect, studied extensively, is believed to play an important role in the structure transformation of different carbon nanostructures, such as the coalescence of fullerenes or the melting of carbon nanotubes.~\cite{Zhao02, Yoon04} According to \textit{ab initio} studies, its energy is of the order of 5--6 eV with an energy barrier energy of the order of 10 eV. \cite{Nardelli02,Blase02,Ma09}

Without the presence of adatoms which can reduce this value within the range 0.7$-$2.3 eV,\cite{Ewels02}  it is then impossible in practice to observe the formation of SW defects at equilibrium. The barrier for healing of an existing SW defect on the other hand is obviously much lower, of the order of 4 eV or less.
\begin{figure}[htbp!]
\begin{center}
\includegraphics[width=0.95\linewidth]{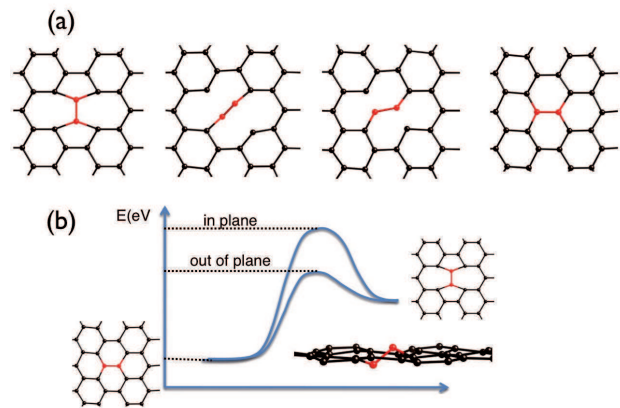}
\end{center}
\caption{(a) Reaction pathway for Stone-Wales transformation observed during a MC simulation at 1000 K. (b) Schematic qualitative representation of the activation barrier for healing the Stone-Wales at high temperatures.}
\label{fig:SW}
\end{figure}
To go further, we use Monte Carlo simulations in the canonical ensemble to investigate the behavior of this defect at high temperatures and without constraining the system. Indeed, many \textit{ab initio} calculations up to now have been performed at 0 K by imposing an in plane Stone Wales $-$ graphene transformation.
In the present work, the system consists of a central defect site within a supercell sufficiently large to minimize boundary effects on the energies of interest. We adopt a structure containing 500 atoms. The full structure is relaxed via standard MC displacements steps. Fig.\ref{fig:SW}(a) shows that when the defect is subjected to high temperatures ($>$ 1000 K) the median bond separating the pentagons and the heptagons completes a 90$^{\circ}$ rotation which results in a perfect $sp^{2}$ hexagonal geometry. During the simulation, we have observed that this thermally activated healing process involves an out-of plane path (Fig.\ref{fig:SW}(b)) (see also Ref.[\onlinecite{Ma09}]). This reaction pathway seems therefore to reduce significantly the activation barrier. \\

\section{Defective graphene}

\begin{figure}[htbp!]
\begin{center}
\includegraphics[width=0.95\linewidth]{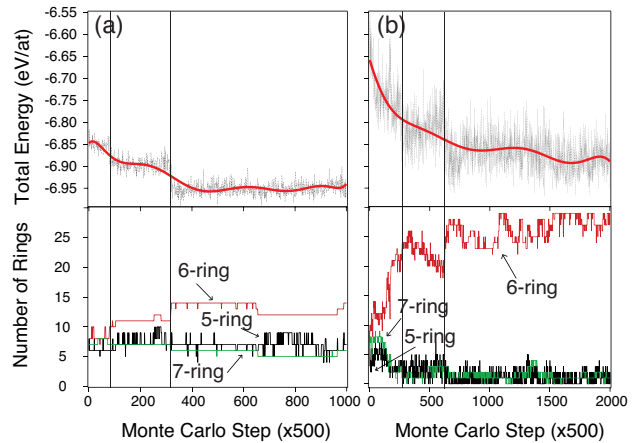}
\end{center}
\caption{Comparison of the total internal energy with the variation of the number of polygons as function of Monte Carlo steps in an isolated graphene sheet at (a) 1000 K and (b) 2500 K.}
\label{fig:steps}
\end{figure}

\begin{figure}[htbp!]
\begin{center}
\includegraphics[width=0.99\linewidth]{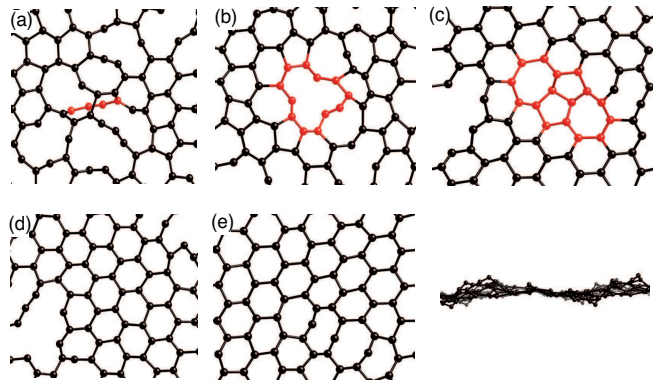}
\end{center}
\caption{Evolution of the defective graphene sheet at 2500 K. The transformation of several defects in an isolated graphene sheet. (a) Linear chain of C helps the transformation of various defects such as vacancies and large polygons. (b) Vacancy defect. (c)-(d) Stone-Wales defect transformation during heating. (e) Final configuration (top and side view)}
\label{fig:2500K}
\end{figure}

We now consider a more complicated situation. A graphene sheet grown by previous GCMC simulations~\cite{Amara02} is submitted to temperatures ranging from 1000 to 2500 K. The initial configuration contains 72 carbon atoms where 70\% of the sheet is populated with defects (see Table \ref{Polygons}). The convergence of the total energy as a function of Monte Carlo steps is controlled and compared with variation of the number of rings in the system. The simulation is stopped when the total energy no longer varies on the average, which implies that the system has reached a Gibbs energy minimum.
\begin{table}[htbp]
\caption{Number of rings in initial and final configurations for different temperatures.}
\begin{tabular}{lccccc}
 \hline
              & 5-ring & 6-ring &  7-ring & 8-ring & 9-ring \\ 
  \hline
Initial     &  6       & 8         & 7         &  5       & 3           \\
1000 K  &  7       & 14       & 6         &  5       & 1           \\
1500 K  &  7       & 11       & 7         &  1       & 1           \\
2000 K  &  3       & 16       & 6         &  2       & 2           \\
2500 K  &  0       & 29       & 1         &  2       & 0           \\
 \hline
\end{tabular}
\label{Polygons}
\end{table}
\begin{figure}[htbp!]
\begin{center}
\includegraphics[width=0.99\linewidth]{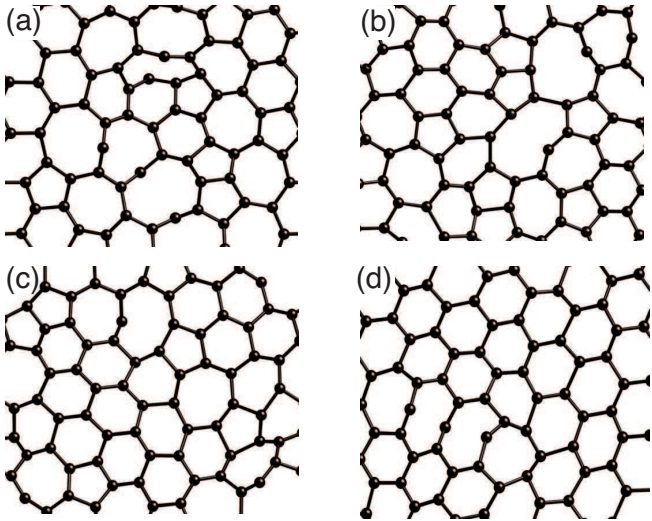}
\end{center}
\caption{Equilibrium configurations at (a) 1000 K, (b) 1500 K, (c) 2000 K and (d) 2500 K.}
\label{fig:pictorialview}
\end{figure}
Fig.\ref{fig:pictorialview} shows the final states of the MC runs for four temperatures ($T=$ 1000, 1500, 2000 and 2500 K). Table \ref{Polygons} presents the corresponding types of polygons. At 1000 K, a significant amount of defects remains in the final configuration. Upon heating, octagons and nonagons are the first to be corrected, implying that their activation barrier is the lowest. Pentagons and heptagons are the most difficult to heal and remain in large concentration until 2500~K. Consequently, topological defects of different sizes are thermally activated at different temperatures. During the simulation at 1000 K, we observe a stepwise decrease of the total internal energy (see Fig.\ref{fig:steps}(a)). The steps coincide with a significant increase in the number of hexagons.  Indeed, the configuration that is favored by the carbon atoms is a three-fold environment, where the bond lengths and angles of an $sp^{2}$ configuration are respected as in the hexagon. Such a geometry greatly stabilizes the system. This tendency is further confirmed by calculations done at 1500, 2000 and 2500 K as  shown in Fig.\ref{fig:steps}(b). However, the steps are less visible at high temperatures ($>$ 1000K) due to the large fluctuations of the system which results into noise in the plots. As expected, the presence of hexagons greatly stabilizes the system and tends to heal all other types of defects. As seen in Fig.\ref{fig:2500K}, where the evolution of the system at 2500 K is presented, we observe a predominance of hexagons at the end of the simulation. The final configuration, corresponding to a carbon structure with 29 hexagons, is indeed an almost perfect graphene sheet.
\section{Role of  nickel}

\begin{figure}[htbp!]
\begin{center}
\includegraphics[width=0.99\linewidth]{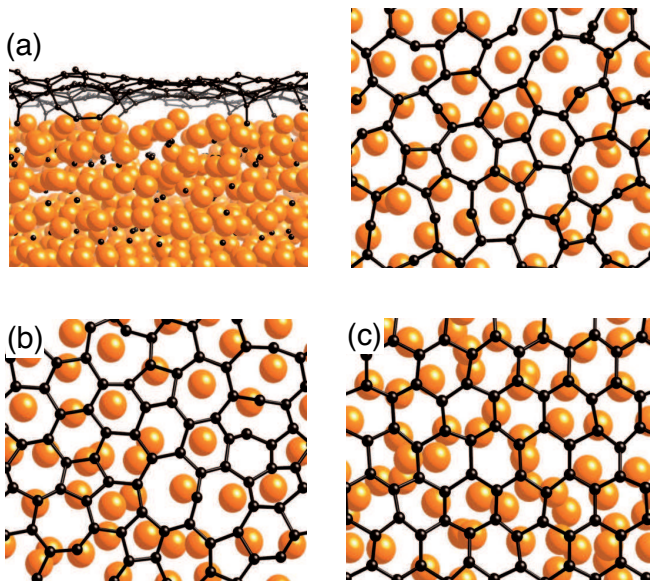}
\end{center}
\caption{Graphene sheet in the presence of the Ni lattice. (a) Side and top views of the initial configurations. Equilibrium configurations at (c) 1000 K and (d) 2500 K.}
\label{fig:NiC}
\end{figure}

\begin{figure}[htbp!]
\begin{center}
\includegraphics[width=0.99\linewidth]{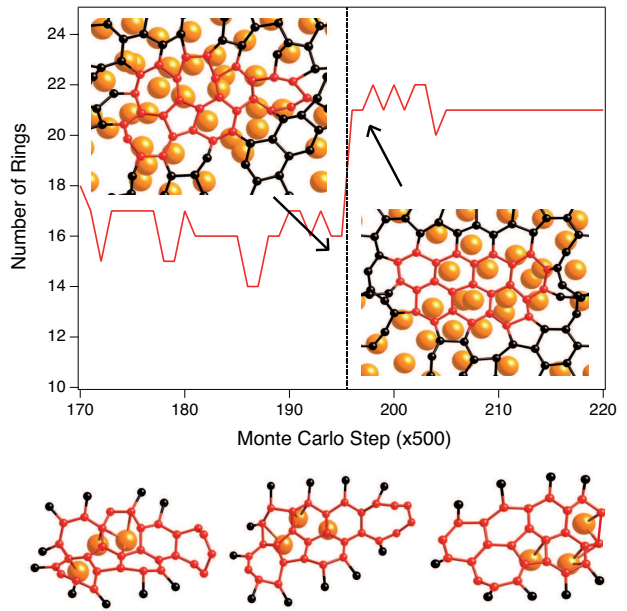}
\end{center}
\caption{Variation of the number of polygons as a function of the MC steps at 2500 K . Ni helping the correction of defects corresponding to a step-like stabilization of the system. We observe that the defective zone (in red) vanishes. This is mainly due to the Ni atoms which interact with this area as seen on the
snapshot.}
\label{fig:Nisurdefaut}
\end{figure}

We further investigate the correction and migration of defects in $sp^{2}$ structures by adding to the previous system the Ni(111) lattice from which the graphene has nucleated (Fig.\ref{fig:NiC}(a) and (b)). We therefore seek to understand the role played by the metal in the healing of defective carbon structures. The NiC system, containing 216 Ni atoms, was subjected to high temperatures ranging from 1000 to 2500 K.
As in the case of the isolated graphene sheet, we notice that at 1000 K, the sheet is far from being healed and a non negligible concentration of defects remains (Fig.\ref{fig:NiC}(c)). Once again, large rings are healed at 1000 and 1500 K whereas pentagons and heptagons cancel out at higher temperatures. At 2500 K the graphene sheet is completely healed as seen in Fig.\ref{fig:NiC}(d). This suggests that the healing of defects is thermally activated as observed for the isolated graphene sheet. By comparison with the case of graphene sheet without catalyst, we deduce that Ni plays a crucial role in defect migration. First, the Ni lattice has contributed to a more rapid stabilization of the system: the graphene sheet was corrected in less Monte Carlo steps. At 1000 K, for an isolated graphene sheet, 2000 MC steps per atom were necessary to form 14 hexagonal rings whereas in presence of nickel atoms the same healing is obtained with only 600 MC steps per atom. Then, the pentagons and heptagons, which were extremely difficult to heal at low temperatures become more reactive in the presence of Ni. To understand the role played by Ni atoms in the healing of defects, we have investigated the step-like increase during the variation of the number of polygons in  a simulation at 2500 K (Fig.\ref{fig:Nisurdefaut}). This detailed analysis suggests that Ni atoms can interact with the defects of the graphene sheet. During the simulation, a defective zone is identified: the red one in Fig.\ref{fig:Nisurdefaut}. The Ni atoms are seen interacting precisely with the defective parts of the graphene sheet whereas no Ni atoms are found in the proximity of the areas of the graphene sheet where a hexagonal geometry has been established. This may not be so surprising; defects change essentially not only the electronic properties but also the chemical properties of carbon structures, being centers of their chemical activities:~\cite{Boukh08} several \textit{ab initio} calculations report on the strong interaction of transition metal impurities with defects.~\cite{Boukh09}  We further observe during the simulation that the interacting Ni atoms catalyze certain reconstructions. The step-like increase in the number of polygons corresponds to Ni atoms interacting with defective areas which help to establish about six hexagonal rings (Fig.\ref{fig:Nisurdefaut}(b)). Once again, these observations are in good agreement with \textit{ab initio} calculations demonstrating that Stone-Wales transformation occurring during growth are made easier by the presence of additional metal atoms, reducing the barrier from about 6 eV to 2.80$-$3.5 eV.~\cite{Charlier07, Jin08}  Moreover, recent \textit{in situ} X-ray diffraction measurements of amorphous carbon on Ni surfaces have suggested graphitization by direct metal-induced crystallisation, confirming the key role played by the metallic substrate to heal defected carbon structures.~\cite{Saenger10}  

\section{Conclusion}

In conclusion, based on tight-binding Monte Carlo simulations, we have studied the healing processes of carbon structures. High temperatures are capable of healing the Stone-Wales defect by adopting an out-of-plane path that significantly reduces its activation barrier. A study of an isolated graphene sheet has demonstrated that defects were thermally activated and that high temperatures are able to heal the structure. Once this sheet is put in contact with the Ni lattice, the defects are healed at a faster rate and at lower temperatures. We have also observed, during our simulations, that the Ni particles participate actively in the healing process. Therefore the presence of Ni favors not only the nucleation of the graphitic structures, but also the healing of defects.

Finally, in a more general context the results reported here are an illustration that defects in $sp^{2}$ carbon structures are highly sensitive to the temperature and to the presence of a transition metal. The present study of graphene may give a key to understanding the properties of other carbon nanostructures, such as carbon nanotubes. Up to now, atomic scale simulations are incapable of explaining the chiral-selective growth observed experimentally.~\cite{Haru09, Chiang09}  The approach proposed here could help to identify individual healing mechanisms during growth that produces perfect tube structures and those favoring a definite chirality.

\begin{acknowledgments}

The authors thank the GDRI Graphene-Nanotubes and the ANR SOS Nanotubes for financial support.

\end{acknowledgments}

\begin{thebibliography}{99}
\bibitem{Jan10}
The authors thank Dr J.H. Los for his contribution to the implementation of these improvements. A detailed account will be published elsewhere.
\end{thebibliography}
\end{document}